\documentclass[aip,graphicx]{revtex4-1}
\usepackage{amsmath, amsfonts}
\usepackage{graphicx}
\usepackage{epstopdf}
\usepackage{color}
\draft % marks overfull lines with a black rule on the right

\begin{document}

\title{Local dissipation limits the dynamics of impacting droplets on smooth and rough substrates} %Title of paper
\author{Yuli Wang}
\email[]{yuli@mech.kth.se}
\affiliation{Department of Mechanics \\ The Royal Institute of Technology, 100 44 Stockholm, Sweden}
\affiliation{Department of Mathematics \\ University of Oslo, 0851 Oslo, Norway}
\author{Gustav Amberg}
\email[]{gustav.amberg@sh.se}
\affiliation{S{\"o}dert{\"o}rn University, 14189 Huddinge, Sweden}
\affiliation{Department of Mechanics \\ The Royal Institute of Technology, 100 44 Stockholm, Sweden}
\author{Andreas Carlson}
\email[]{acarlson@math.uio.no}
\affiliation{Department of Mathematics \\ University of Oslo, 0851 Oslo, Norway}

\date{\today}

\begin{abstract}
A droplet that impacts onto a solid substrate deforms in a complex dynamics. To extract the principal mechanisms that dominate this dynamics we deploy numerical simulations based on the phase field method. Direct comparison with experiments suggests that a dissipation local to the contact line limits the droplet spreading dynamics and its scaled maximum spreading radius $\beta_\mathrm{max}$. By assuming linear response through a drag force at the contact line, our simulations rationalize experimental observations for droplet impact on both smooth and rough substrates, measured through a single contact line friction parameter $\mu_f$. Moreover, our analysis shows that at low and intermediate impact speeds dissipation at the contact line limits the dynamics and we describe $\beta_\mathrm{max}$ by the scaling law $\beta_\mathrm{max} \sim (Re \mu_\mathrm{l}/\mu_f)^{1/2}$ that is a function of the droplet viscosity ($\mu_l$) and its Reynolds number ($Re$).
\end{abstract}


\maketitle %\maketitle must follow title, authors, abstract and \pacs

\section{Introduction}
Impact of liquid droplets onto a solid substrate is essential to applications such as spray coating \cite{dykhuizen1994review}, ink-jet printing\cite{attinger2000experimental}, additive manufacturing \cite{fathi2010regimes} and pesticide deposition\cite{bergeron2000controlling}. Upon impact with the substrate the droplet deforms in a complex dynamics, where a gas film can become trapped underneath the droplet \cite{mehdizadeh2004formation,mandre2009precursors,duchemin2011curvature,kolinski2012skating,bouwhuis2012maximal,liu2013compressible,visser2015dynamics,li2015time} and as it spreads create a splash by droplet ejection at the tip of its spreading front\cite{driscoll2011ultrafast,yokoi2011numerical,visser2012microdroplet,riboux2014experiments,xu2005drop}. The droplet deformation and spreading is typically driven by its inertia and hindered by viscous and surface tension forces. Two non-dimensional numbers are particularly relevant to characterize the dynamics, which is the Reynolds number $Re = \rho_\mathrm{l}V_i 2R / \mu_\mathrm{l}$ giving the ratio between inertia and viscous forces and the Weber number $We = \rho_\mathrm{l}V_i^2 2R / \sigma $ which gives the ratio between inertia and surface tension forces. $\sigma$ is the surface tension coefficient of the gas-liquid, $\rho_\mathrm{l}$ is the liquid density, $\mu_\mathrm{l}$ is the liquid viscosity and $V_\mathrm{i}$ is the droplet impact speed. Besides inertia, viscosity and surface tension, we hypothesize and show that a dissipation local to the contact line can limit the droplet dynamics on both smooth and rough substrates.

\begin{figure}[h]
 \centering
    %\begin{align}
     \includegraphics[height = 0.4\textwidth,trim = 0cm 0cm 0cm 0cm, clip=true]{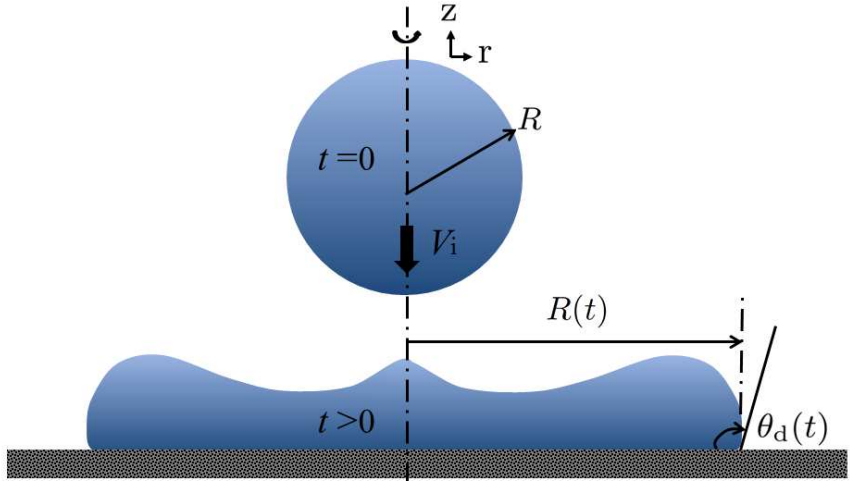}
      \caption{\label{f1} Sketch of the axi-symmetric computational domain and the droplets initial condition with $R$ the initial droplet radius and $V_i$ the impact speed. The spreading radius $R(t)$ and the apparent dynamic contact angle $\theta_\mathrm{d}(t)$ is illustrated as the droplet has spread onto the solid substrate.}
      \end{figure}

One parameter that describes the droplet impact dynamics and is typically quantified is the spreading factor $\beta(t) = R(t)/R$, where $R(t)$ is the droplet spreading radius, $R$ is the initial droplet radius and $\beta_\mathrm{max}=\max(\beta(t))$, see Fig.\ref{f1}. Two primary regimes have been identified to describe $\beta_\mathrm{max}$; an inertia-viscous regime where $\beta_\mathrm{max} \sim Re ^{\frac{1}{5}}$ \cite{fedorchenko2005effect,roisman2002normal} and an inertia-capillary regime where $\beta_\mathrm{max} \sim We ^{\frac{1}{2}}$ \cite{bennett1993splat, lagubeau2012spreading}. A single law has been derived to connect these two regimes $\beta_\mathrm{max} Re^{-\frac{1}{5}} \sim f(WeRe^{-\frac{2}{5}})$ \cite{eggers2010drop}, which has rationalized experiments for a wide range of $Re$ and $We$ numbers \cite{laan2014maximum}. Other scaling laws for $\beta_\mathrm{max}$ with different exponents for $Re$ and $We$ have been proposed \cite{scheller1995newtonian,pasandideh1996capillary, mao1997spread,chandra1991collision,vadillo2009dynamic,ukiwe2005maximum,clanet2004maximal}, which include additional effects such as the substrate wettability. A detailed description of these different scaling laws can be found in the recent review by Josserand and Thoroddsen\cite{josserand2016drop}. However, none of these scaling laws describe $\beta_\mathrm{max} $ at the low impact speeds as they predict $\beta_\mathrm{max} \rightarrow 0$ as $V_\mathrm{i} \rightarrow 0$, which is not true for any case with an equilibrium contact angle $\theta_\mathrm{e} < 180 ^\circ$. To mitigate this artifact the maximum spreading radius $\beta_0$ for $V_\mathrm{i} = 0$ has been included into the analysis $\sqrt{\beta_\mathrm{max}^2 - \beta_0^2} = Re^{\frac{1}{5}}We^{\frac{1}{2}}/(A + We^{\frac{1}{2}})$ that agrees favorably with experimental data for both low and high impact speeds\cite{lee2015universal}, where $A$ is an ad-hoc fitting parameter.

Substrate roughness is another parameter that can influence the droplet impact dynamics\cite{kannan2008drop,vaikuntanathan2016maximum,lee2016modeling,tsai2009drop,van2014microstructures}. Droplet impact on regular micro-textured substrates\cite{kannan2008drop,vaikuntanathan2016maximum,robson2016asymmetries} show that $\beta_\mathrm{max}$ is influenced by the substrate topography. Even a substrate with small aspect ratio roughness hinders droplet spreading \cite{lee2016modeling}, although the effect becomes less pronounced. %texture of the solid substrate can make the droplet dynamics deviate from the scaling laws that work well for smooth substrates.

In this work we focus on describing $\beta(t)$ and $\beta_\mathrm{max}$ in the regime of non-splashing droplets\cite{yarin2006drop,josserand2016drop} i.e. small and intermediate impact speed. We show that as in a spontaneous droplet spreading process\cite{carlson2011dissipation,carlson2012contact}, a detailed description of the physical processes at the contact line must be included to accurately describe the interface dynamics. Numerical experiments based on the phase field method and the Navier Stokes equations show that friction local to the contact line limits the dynamics and generates a significant dissipation. We treat the contact line friction parameter $\mu_f$ as a material property for each combination of air-liquid-solid, which should be independent of the impact speed. We determine the magnitude of $\mu_f$ by directly comparing the numerical simulations with several independent experiments \citep{pasandideh1996capillary,laan2014maximum,lee2015universal,lee2016modeling,vaikuntanathan2016maximum}. Our assumption of linear response through a Stokes-like drag at the contact line shows that the simulations can accurately reproduce experimental observations. We further extend our analysis to rough substrates and rationalize the differences in the dynamics compared with smooth substrates. Finally, we show that the regime where the principal dissipation is local to the contact line is described by a scaling law based on the contact line friction parameter $\mu_f$.
%In order to group different types of surfaces into a single framework, a universal treatment accounting for liquid-solid iteration at the surface is still missing. This treatment is also necessary for capturing the dynamic wetting process which becomes important for predicting $\beta_\mathrm{max}$ at the low asymptotic $V_\mathrm{i}\rightarrow 0$. We here propose the contact line friction, a local dissipative mechanism which occurs at a moving contact line, to play this role. This mechanism is firstly applied to the spontaneous spreading of droplets \cite{carlson2011dissipation,carlson2012contact}, and is extended its usage for droplet impact in this work. We prove through numerical simulations that by using a single contact line friction parameter $\mu_f$, one can accurately predict $\beta_\mathrm{max}$ on smooth and rough surfaces for low and high impact speeds.

\subsection{Models and Methods}
%\paragraph{This is the next level heading.~~} For this level please use \texttt{\textbackslash paragraph}. These headings should also end in a full point.
We describe the multiphase system by using the phase field method \cite{jacqmin1999calculation}, which considers the two binary phases (gas liquid) as a mixture. The mathematical model is composed of the Cahn-Hilliard equation \cite{cahn1958free} Eq.(\ref{eq1},\ref{eq2}), which is coupled with the Naiver-Stokes equations Eq.(\ref{eq3},\ref{eq4}) for an incompressible fluid flow\cite{jacqmin1999calculation};

\begin{equation}
\frac{\partial C}{\partial t} + \mathbf{u} \cdot \nabla C = \gamma \nabla ^2 \phi
\label{eq1}
\end{equation}
\begin{equation}
\phi = -\frac{3}{2\sqrt{2}} \sigma \left( \epsilon \nabla^2 C  - \frac{C^3-C}{\epsilon} \right)
\label{eq2}
\end{equation}
\begin{equation}
\rho(C) \left( \frac{\partial \mathbf{u}}{\partial t} + (\mathbf{u} \cdot \nabla) \mathbf{u} \right) = -\nabla P +  \nabla \cdot (\mu(C)(\nabla \textbf{u} + (\nabla \mathbf{u}))^T) + \phi \nabla C
\label{eq3}
\end{equation}
\begin{equation}
 \nabla \cdot \textbf{u} = 0.
 \label{eq4}
\end{equation}
$C=C(r,z,t)$ is an order parameter that varies smoothly from $C=-1$ (gas) to $C=1$ (liquid) between the two immiscible phases and $\left|C\right|<1$ indicates that the interfacial region that has a finite thickness $\epsilon$. $\phi=\phi(r,z,t)=\delta F(r,z,t)/\delta C$ is the chemical potential, given by the variation of the systems postulated free energy $F(r,z,t)$ that has an interfacial and bulk free energy term. The free energy is required to reduce with time i.e. $\gamma>0$ and $\gamma$ is the mobility factor that controls the interfacial diffusion. $\mathbf{u} = \mathbf{u}(r,z,t)$ is the velocity, $P = P(r,z,t)$ is the pressure, whereas the density $\rho(C)=(1+C)\rho_\mathrm{l}/2 + (1-C)\rho_\mathrm{g}/2$ and the viscosity $\mu(C)=(1+C)\mu_\mathrm{l}/2 + (1-C)\mu_\mathrm{g}/2$ are interpreted as function of $C$. The air surrounding the droplet is assumed at atmospheric pressure with a density $\rho_\mathrm{g} = 1.23 \mathrm{kg/m^3}$ and a viscosity $\mu_\mathrm{g} = 1.81 \times 10^{-5} \mathrm{Pa.s}$. The material properties of the droplet, along with the impact speeds, equilibrium contact angles and range of simulated $Re$ numbers and $We$ numbers are listed in Table \ref{tab1}.

All simulations are performed with a no-slip boundary condition for the velocities at the solid substrate ($\mathbf{u} = 0$) and all other boundaries are assumed to be in contact with ambient air at constant pressure ($P=0$) and with no-flux of the chemical potential ($\nabla \phi\cdot \mathbf{n}=0$ with $\mathbf{n}$ as the boundary normal). To model the contact line dynamics we use the non-equilibrium boundary condition\cite{jacqmin2000contact},
\begin{equation}\label{eq5}
\frac{2\sqrt{2}}{3} \sigma \epsilon  \nabla C \cdot \mathrm{n} + \sigma \mathrm{cos}(\theta_\mathrm{e})g'(C) =  -\mu_f\epsilon \frac{\partial C}{\partial t}
\end{equation}
where $\mu_f$ is interpreted as a friction factor at the contact line and $g(C) = -2/4-3/4C+1/4C^3$ provides a transition for the dry (gas-solid) or wet (liquid-solid) substrate surface tension. For $\mu_f = 0$ we assume local equilibrium at the contact line and the interface adopts the equilibrium contact angle $\theta_\mathrm{e}$, but if $\mu_f>0$ the contact angle becomes dynamic and allowed to deviate from $\theta_\mathrm{e}$.

We use the following scaling; $\mu^*(C) = \mu(C)/\mu_\mathrm{l}, \rho^*(C)=\rho(C)/\rho_\mathrm{l}, \mathbf{u}^* = \mathbf{u}/V_\mathrm{i}, t^* = tV_\mathrm{i}/R, P^* = PR/(\mu_\mathrm{l}V_\mathrm{i}), \phi^* = \phi2\sqrt{2}\epsilon/ (3\sigma)$ to make Eq.\eqref{eq1}-Eq.\eqref{eq5} non-dimensional, where the superscript $*$ denotes non-dimensional variables. In addition to the Reynolds number $Re$ four non-dimensional numbers appear in the equations; the Capillary number $Ca= We/Re = V_\mathrm{i} \mu_\mathrm{l} / \sigma$ gives the ratio of the viscous force to the surface tension force, the non-dimensional friction parameter $D_\mathrm{w} = (\epsilon \mu_f V_\mathrm{i})/(R \sigma)$ gives the ratio of the contact line friction force to the surface tension force, the Cahn number $Cn=\epsilon / R = 0.005$ gives the ratio of the interface thickness and the droplet radius, and the P\'eclet number $Pe = 2\sqrt{2}V_\mathrm{i}\epsilon R/(3\sigma \gamma) = 100$ gives the ratio of advection to diffusion. Both $Cn$ and $Pe$ are fixed in all of our simulations such that the results satisfy the sharp interface criterion\citep{yue2010sharp,magaletti2013sharp}, $\epsilon = 0.005R$ and $\gamma = (2\sqrt{2}V_\mathrm{i}\epsilon R)/(300\sigma)$. The contour line $C=0$ is interpreted as the droplet interface and used to extract $\beta(t),\beta_\mathrm{max}$ and $\theta_\mathrm{d}(t)$.

The numerical simulations are performed with FemLego\citep{amberg1999finite}, a symbolic finite element toolbox that solves partial differential equations. All simulations are performed in an axi-symmetric coordinate system where the domain extends $10R$ in the $r$ direction and $5R$ in the $z$ direction. An adaptive mesh refinement method is used to enable a high resolution of the interface with a minimum mesh size of $\Delta r \approx \Delta z\approx 0.001R$, which resolves the interface with $Cn/\Delta r\approx5$ cells. The droplet's center of mass is initialized at a height $z=1.5R$ from the solid substrate and the dynamic contact angle $\theta_\mathrm{d}(t)$ is measured at a height of $z=100 \mathrm{\mu m}$ using linear interpolation along $C=0$ similar to the method used to in the experimental analysis that we are directly comparing against\cite{lee2015universal}.

The phase field method has previously been used with success to simulate droplet impact dynamics\citep{khatavkar2007diffuse,zhang2016phase} to quantify the early spreading and bubble entrapment, in accordance with experimental observations\cite{wang2016events}. However, none of these account for a dynamic contact angle treatment in Eq.\eqref{eq5} with $\mu_f>0$ or quantify the maximum spreading radius of the droplet, which we will show are two intertwined processes needed to rationalize the impact dynamics of droplets on smooth and rough solid substrates.

   \begin{table*}
   \small
    \caption{Simulated material properties and droplet impact conditions on substrates with different equilibrium contact angles ($\theta_\mathrm{e}$([steel, stainless steel, aluminium,grooved stainless steel]${_\mathrm{air-water}}$)=[$61^\circ$, $90^\circ$, $94^\circ$, $130^\circ$] and $\theta_\mathrm{e}($[steel]${_\mathrm{air-glycerol/water}}$)=[$52^\circ$]).}
    \label{tab1}
    \begin{tabular*}{\textwidth}{@{\extracolsep{\fill}}lllllllll}
    \hline
     $ $   & $\rho_l(\mathrm{kg\cdot m^{-3}})$ & $\mu_\mathrm{l} (\mathrm{Pa.s})$ & $\sigma (\mathrm{N\cdot m^{-1}}) $ & $R(\mathrm{mm})$   &   $V_i(\mathrm{m/s})$ & $\theta_e (^\circ)$ & $Re$ & $We$ \\
     \hline
     water   & 1000  & 0.001  & 0.073  & 1 & 0.28-4.85   &  61,90,94,130  &320-$10^4$  & 0.6-664\\
     glycerol-water & 1158 & 0.01 & 0.068 & 0.92 & 0.19-9.28 & 52 &40-1956  & 1-2653 \\
     \hline
      \end{tabular*}
    \end{table*}

\section{Results and discussions}\label{sec:res}
\subsection{Droplet impact on smooth substrate}
Phase field simulations of droplets that spontaneously spread onto a solid substrate has shown that in order to accurately describe the contact line dynamics a local dissipation by using $\mu_f>0$ in Eq.\ref{eq5} is required\cite{carlson2012contact}. The mathematical form of Eq.\ref{eq5} comes from the assumption of linear response with a reduction of the free energy in time, and can be interpreted as a Stokes-like drag at the contact line. We interpret this wall-interface friction parameter $\mu_f$ as a physical property that depends on the combination of the gas-liquid-solid. We hypothesize that parts of the parameter space that compose the droplet impact dynamics can only be described with an accurate local treatment of the dynamic contact angle through $\mu_f$.

\begin{figure*}
    \centering
    \includegraphics[height = 0.35\textwidth,trim = 0cm 0cm 0cm 0cm, clip=true]{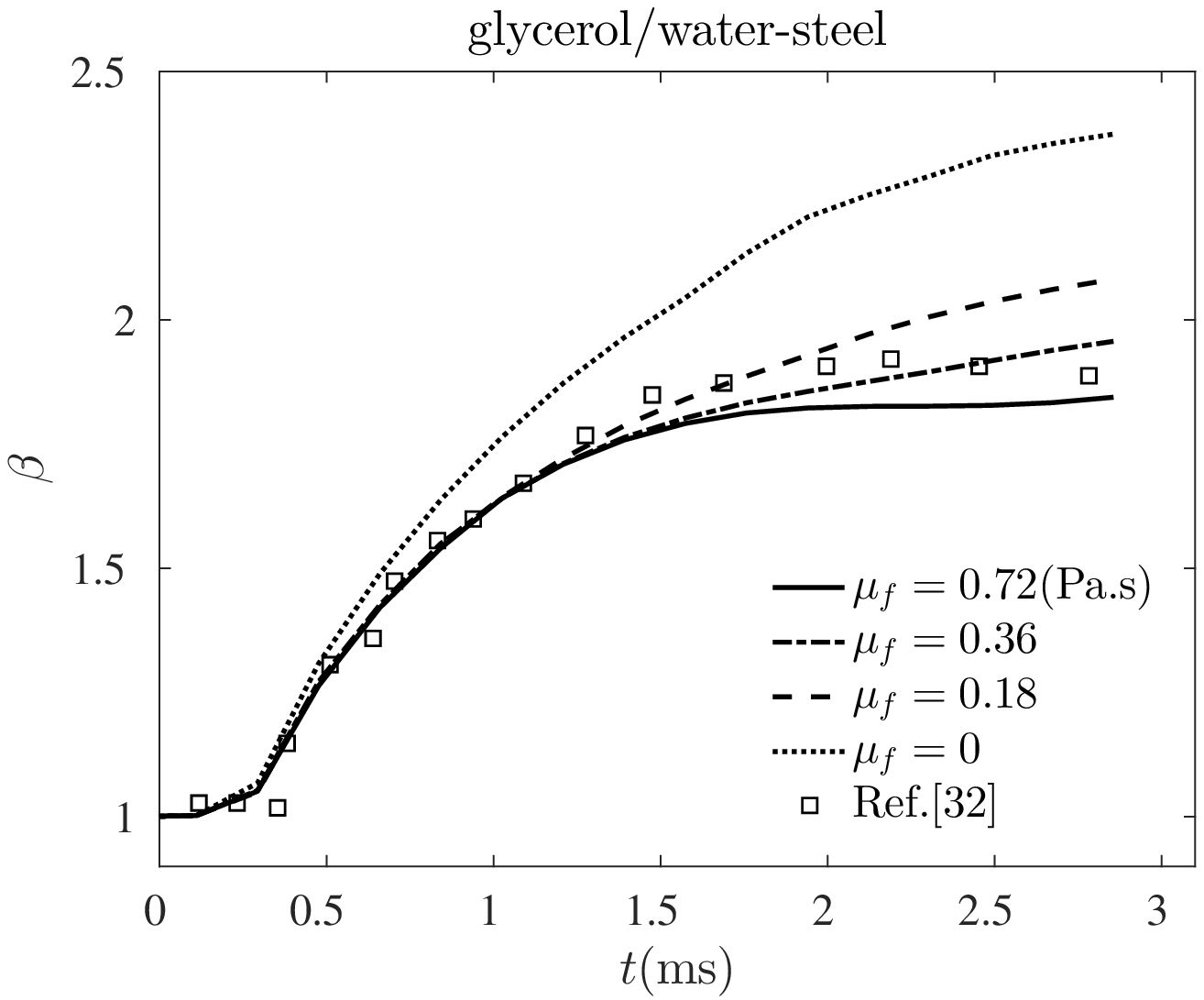}
    \put(-215,175){\small{(a)}}
    \includegraphics[height = 0.35\textwidth,trim = 0cm 0cm 0cm 0cm, clip=true]{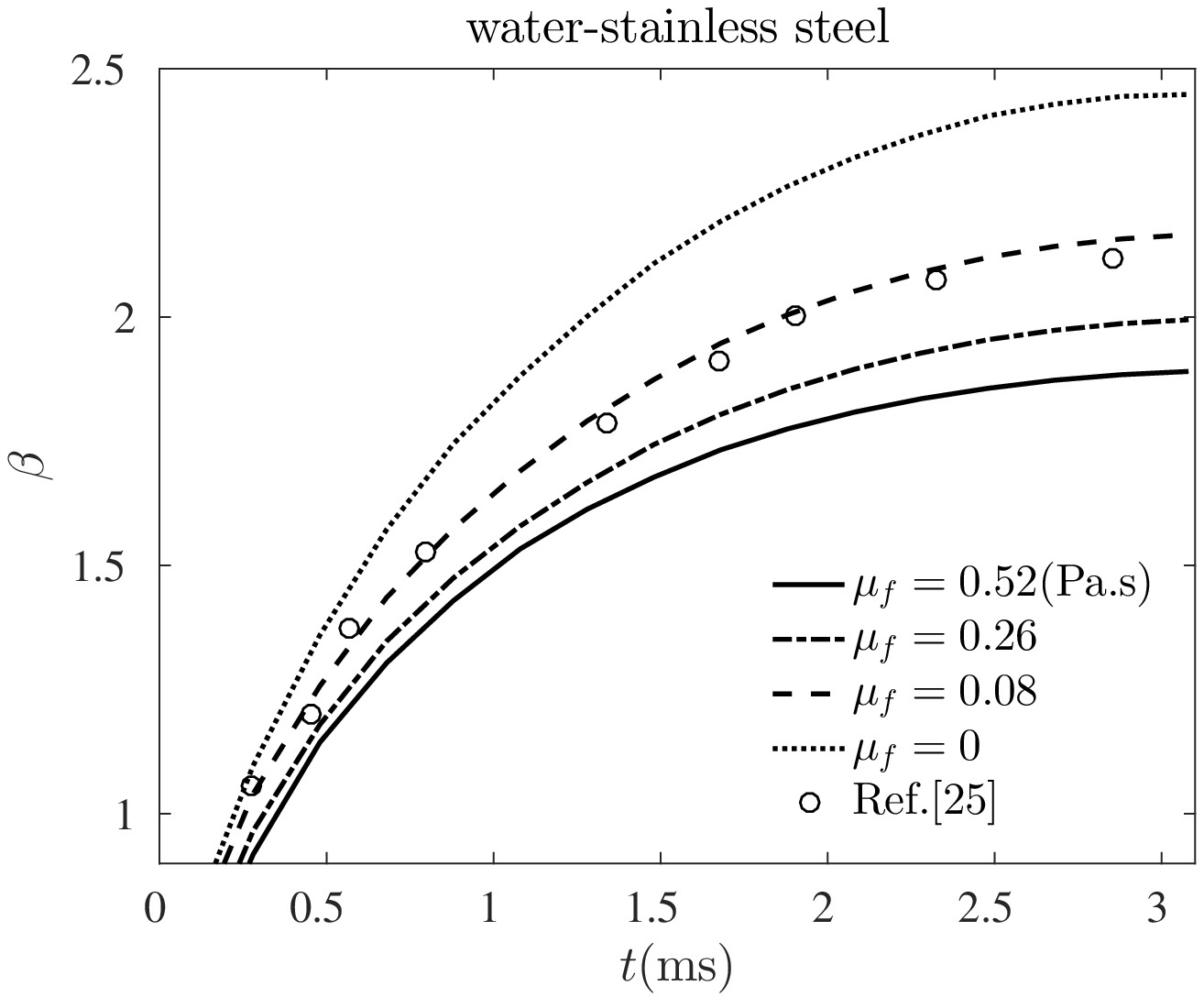}
    \put(-210,175){\small{(b)}}  \\
    \includegraphics[height = 0.35\textwidth,trim = 0cm 0cm 0cm 0cm, clip=true]{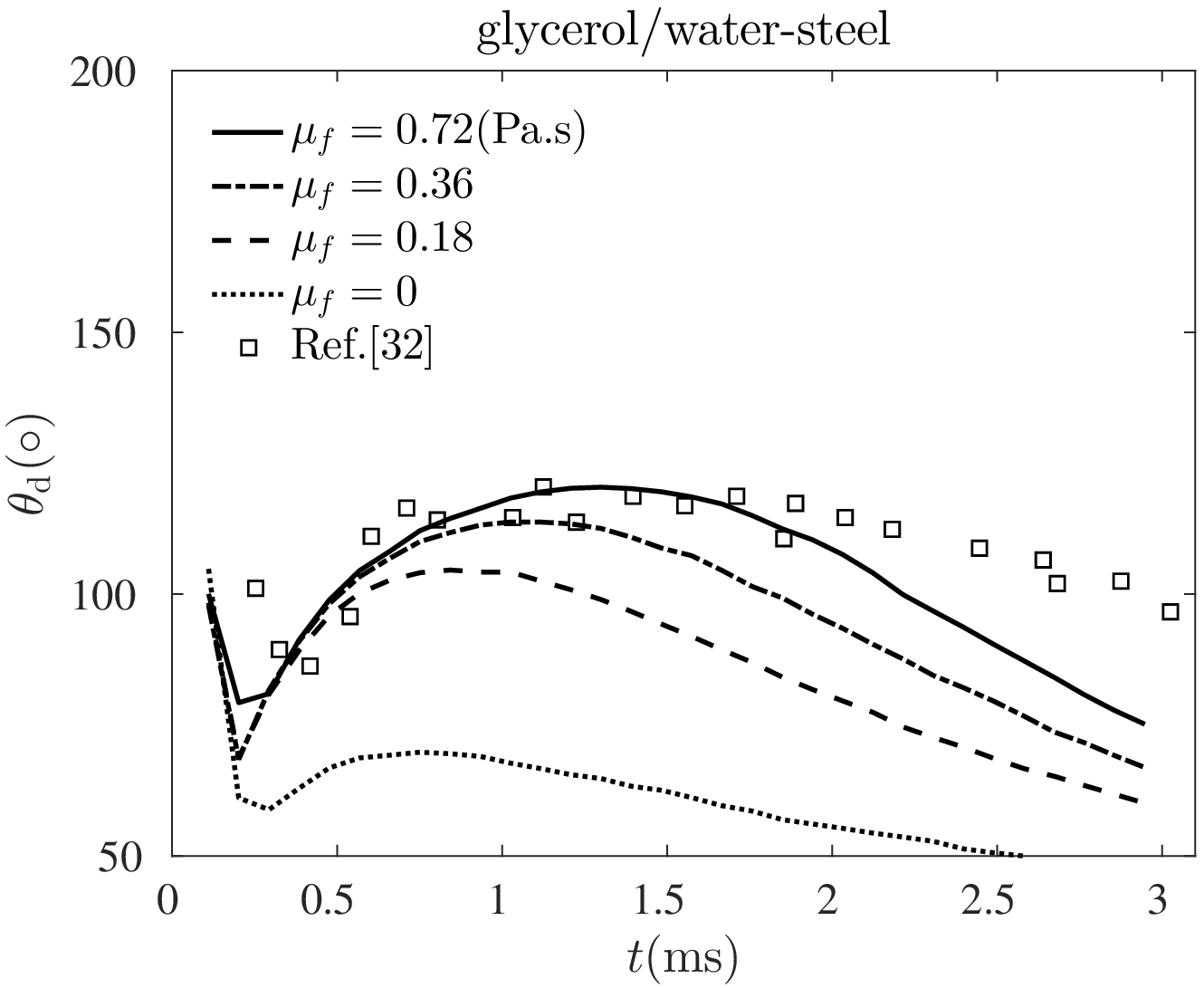}
    \put(-215,175){\small{(c)}}
    \includegraphics[height = 0.35\textwidth,trim = 0cm 0cm 0cm 0cm, clip=true]{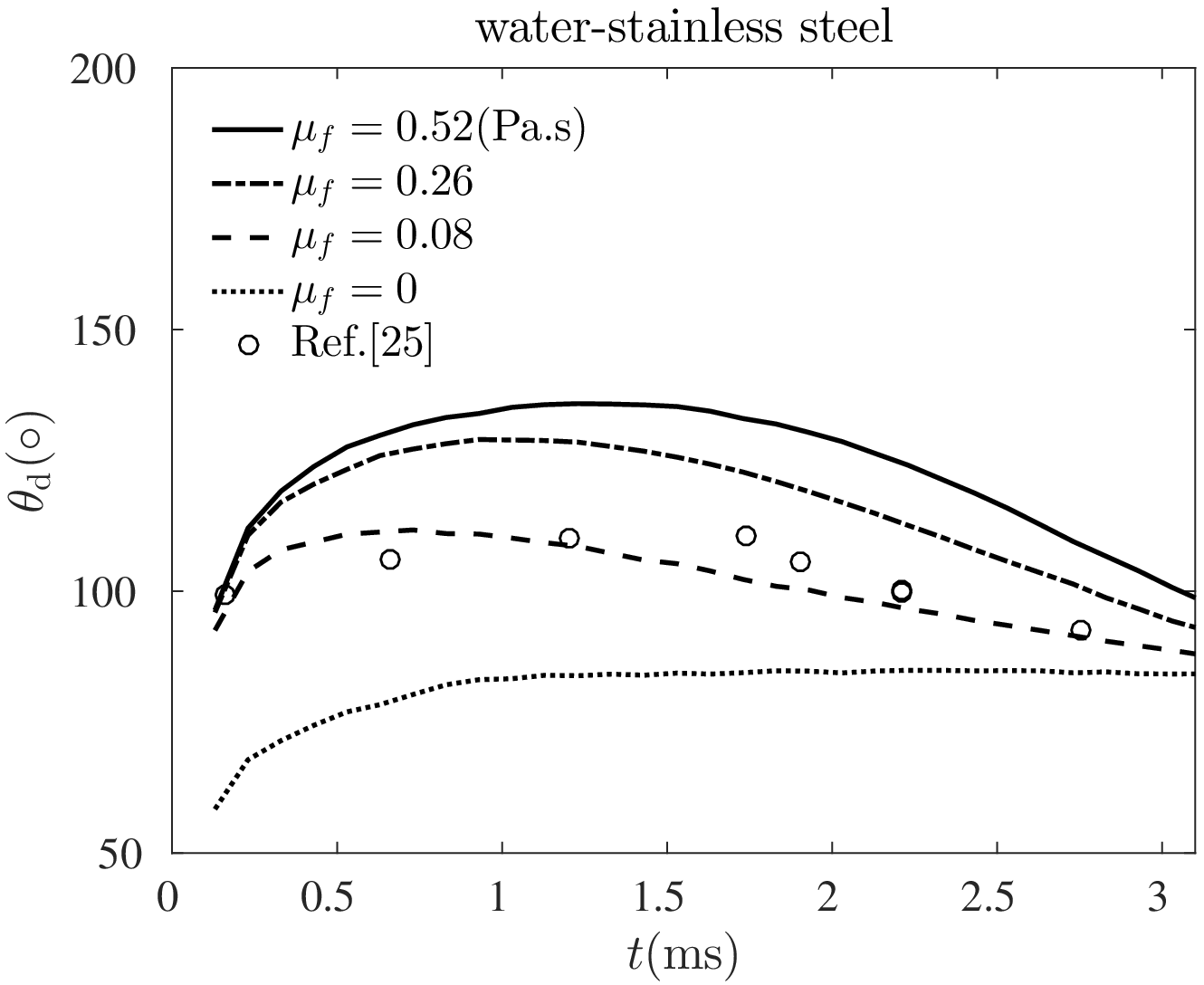}
    \put(-210,175){\small{(d)}}
     \caption{\label{f3}{Comparison of experimental data (markers) and numerical simulations (lines) for a droplet impacting onto smooth substrates with a speed $V_i = 1 \mathrm{m/s}$.  Influence of (a-b) $\beta(t)$ and (c-d) $\theta_\mathrm{d}(t)$ as a function of the viscosity ($\mu_\mathrm{l}$) and the contact line friction parameter ($\mu_f$), using the same definition of $\beta(t)$ as reported in experiments\citep{lee2015universal,lee2016modeling}. }}
    \end{figure*}

Since $\mu_f$ is not known a-priori we determine its magnitude by directly comparing simulations with experiments\citep{pasandideh1996capillary,lee2015universal}, where $\mu_f$ is identified as the best-match with $\beta(t)$ (see Fig.\ref{f3}). Our simulations of droplets of water and glycerol-water mixture show that $\mu_f$ clearly affects the spreading dynamics as well as the shape of the droplet. For water droplets on steel $\mu_f \sim 0.52 \mathrm{Pa.s}$, while increasing its viscosity by introducing glycerol ($\mu_\mathrm{l} = 0.01 \mathrm{Pa.s}$) also increases $\mu_f \sim 0.72 \mathrm{Pa.s}$. These magnitudes for $\mu_f$ are in accordance with previous measurements on spontaneous spreading droplets ($V_i=0$)\cite{carlson2012contact}. Our simulations clearly show that $\mu_f$ controls the time scale for $\theta_\mathrm{d}(t)$ to approach the equilibrium angle, where $\theta_\mathrm{d}(t)$ is the droplets apparent contact angle, see Fig.\ref{f3}(b,d). It is noteworthy that the assumption of local equilibrium, i.e. equilibrium contact angle, over-predicts the spreading factor $ \beta(t)$ and its maximum $\beta_\mathrm{max} = \mathrm{max}( \beta(t))$ for both liquids. Thus, to obtain agreement between the simulations and experiments we need to account for dissipation at the contact line and the local equilibrium assumption fails to capture the spatiotemporal droplet dynamics, see Fig.\ref{f3}(b,d).

\begin{figure}[h]
    \centering
    \includegraphics[height = 0.1\textwidth,trim = 0.0cm 0cm 0cm 0cm, clip=true]{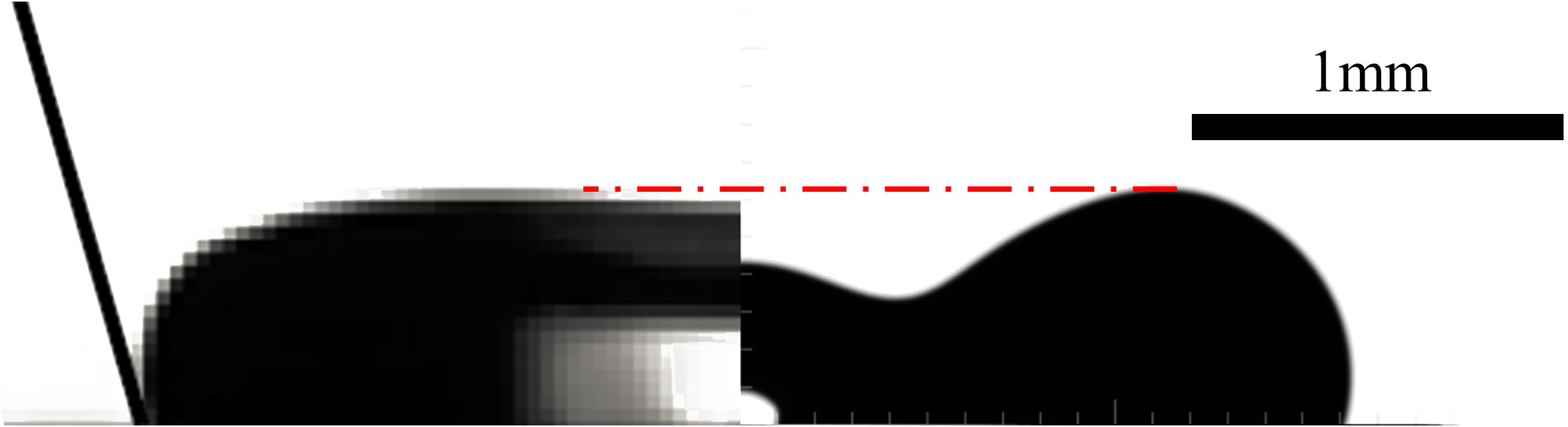}
    \put(-100,45){ $\theta_\mathrm{d} = 101.4^\circ $}
    \put(-170,45){ $\theta_\mathrm{d} = 105.9^\circ $}
    \put(-230,45){ $ \left(a \right) $}\nonumber \\
    \includegraphics[height = 0.1\textwidth,trim = 4.1cm 0cm 4.1cm 0cm, clip=true]{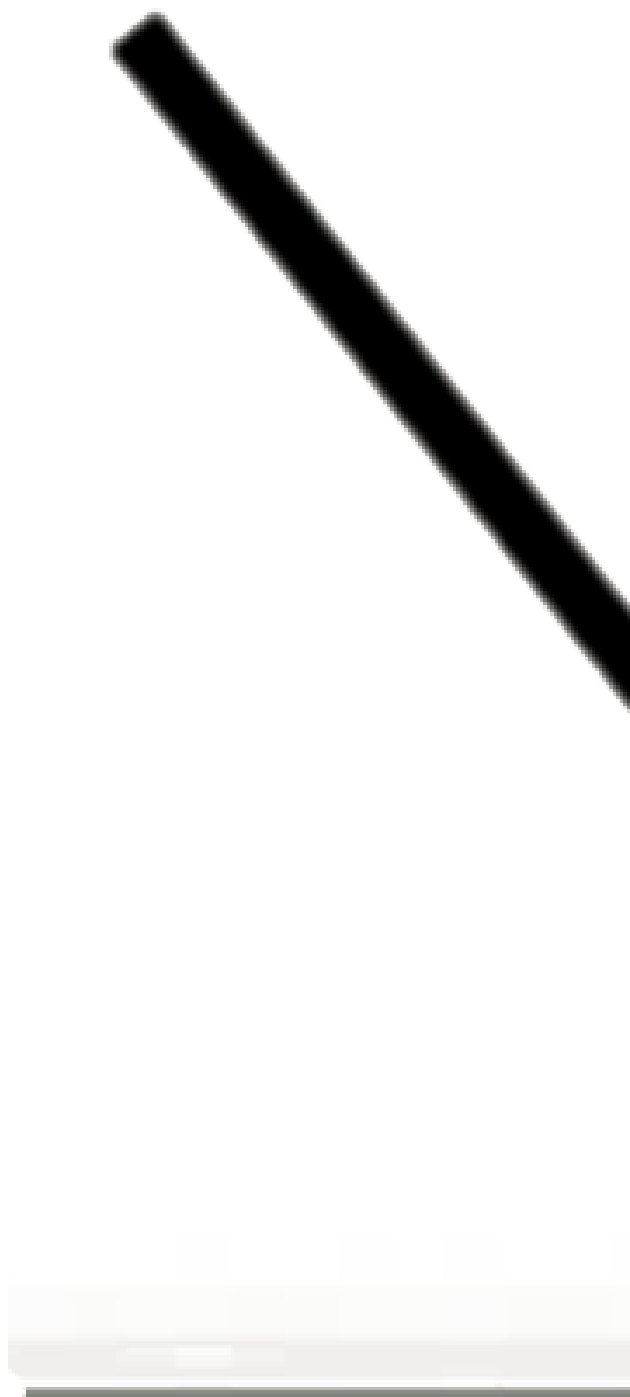}
    \put(-110,45){ $\theta_\mathrm{d} = 126.3^\circ $}
    \put(-180,45){ $\theta_\mathrm{d} = 128.3^\circ $}
    \put(-240,45){ $ \left(b \right)$}\nonumber \\
    \includegraphics[height = 0.1\textwidth,trim = 0.7cm 0cm 0.7cm 0cm, clip=true]{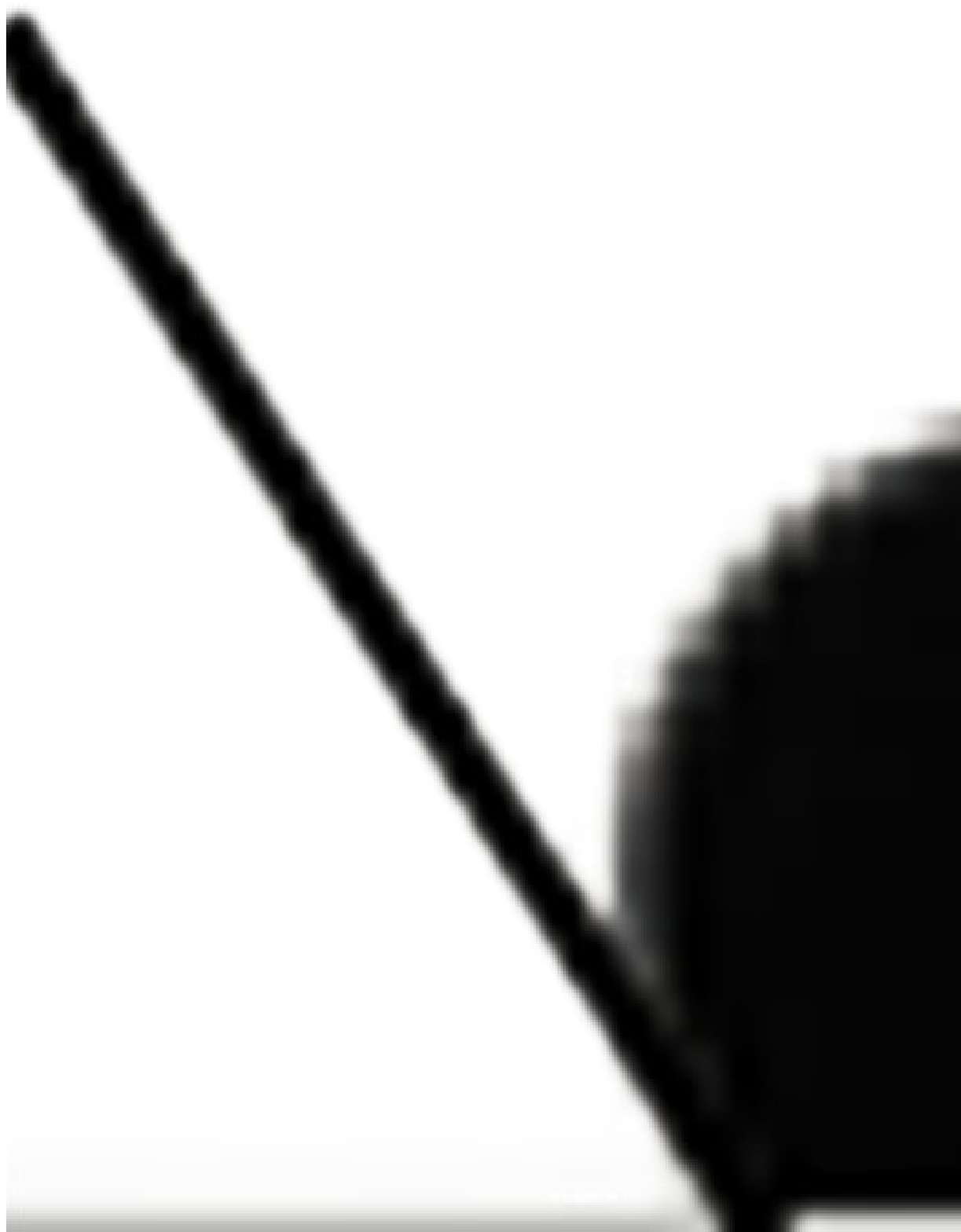}
    \put(-100,40){ $\theta_\mathrm{d} = 114.3^\circ $}
    \put(-170,40){ $\theta_\mathrm{d} = 122.2^\circ $}
    \put(-230,40){ $ \left(c \right)$} \nonumber \\
    \includegraphics[height = 0.1\textwidth,trim = 0cm 0cm 0cm 0cm, clip=true]{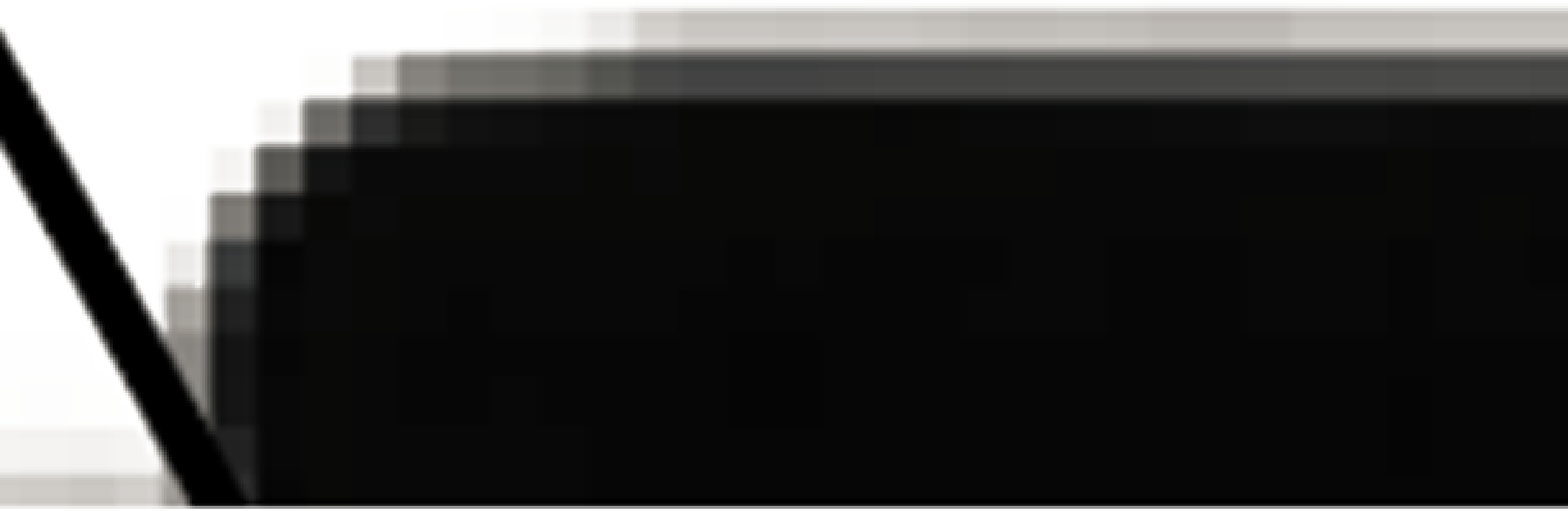}
    \put(-100,40){ $\theta_\mathrm{d} = 114.5^\circ $}
    \put(-170,40){ $\theta_\mathrm{d} = 117.7^\circ $}
    \put(-230,40){ $ \left(d \right)$}\nonumber
    \caption{\label{f2}Comparison of the experimental\citep{lee2015universal} (the left half, their Fig.2) and numerical (the right half) droplet shape as it impacts onto a steel substrate. (a) A water droplet ($1$ mPa.s) with $\mu_f = 0.52 \mathrm{Pa.s}$ and $V_i = 0.57 \mathrm{m/s}$. (b-d) A glycerol-water droplet ($10$ mPa.s) with $\mu_f = 0.72 \mathrm{Pa.s}$ and (b) $V_i = 0.28 \mathrm{m/s}$, (c) $V_i = 0.6 \mathrm{m/s}$, (d) $V_i = 1.86 \mathrm{m/s}$.}
    \end{figure}

After determining $\mu_f$ (Table 2) from an experiment for one impact speed $V_i$, we now assume $\mu_f$ to be a constant material parameter that must be independent of $V_i$.
  \begin{table}
   \small
    \caption{Measurement of $\mu_f$ as a combination of air-liquid-solid combination.}
    \label{tab2}
\centering
    \begin{tabular}{llll}
    \hline
     $liquid-substrate$   & $\theta_e (^\circ)$ & $\mu_f$ (Pa$\cdot$s) & $experiments$\\
     \hline
     glycerol/water-steel& 52$^{\circ}$ & 0.72 & Ref.[32]\\
     water-stainless steel & 90$^{\circ}$ &0.08 & Ref.[25] \\
     water-steel   & 61$^{\circ}$  & 0.52 & Ref.[32]\\
     water-aluminium   & 94$^{\circ}$  &0.08 & Ref.[35]\\
     \hline
      \end{tabular}
    \end{table}
We challenge our hypothesis that $\mu_f$ is unique for a specific air-liquid-solid combination by directly comparing our simulations with experiments \citep{lee2015universal,lee2016modeling} for different $V_\mathrm{i}$. It is clear that as we increase $V_\mathrm{i}$ the droplets aspect ratio i.e. maximum height divided by $\beta_\mathrm{max}$, decreases at $\beta(t)=\beta_\mathrm{max}$. The simulated droplet shapes are in very good agreement with the experiments at $\beta_\mathrm{max}$, where the difference in profiles (Fig.\ref{f2}(a) and (c-d)) is caused by the experimental side-view photos. Since the simulations show a slice through the droplet, they represent its actual shape. The dash-dot lines in panels to the right in Fig.\ref{f2} illustrate how the shape of the droplet would look if we instead would have made a side-view image. Our simulations also capture the entrapment of an air bubble at the symmetry axis at the wall, as seen in Fig.\ref{f2}(a-c).
%For small $V_\mathrm{i} \sim 0.3 \mathrm{m/s} $ the glycerol droplet has an aspect ratio that is close to one, while a spherical cap in Fig.\ref{f2}(b); for larger $V \sim 2 \mathrm{m/s}$ the droplet forms a pancake shape with its spreading tip lifting up, see in Fig.\ref{f2}(d). The droplet height measured at the position on the profile with a maximum value of $z$ , the wetting radius, and $\theta_\mathrm{d}$ all agree well with the experiments, applied the right values of $\mu_f$ for the specified liquids and surfaces. For all cases in Fig.\ref{f2} $\theta_\mathrm{d}$ is much larger than the prescribed $\theta_\mathrm{e}$. Note the detergency

\begin{figure}[h]
    \centering
    \includegraphics[width = 0.55\textwidth,trim = 0cm 0cm 0cm 0cm, clip=true]{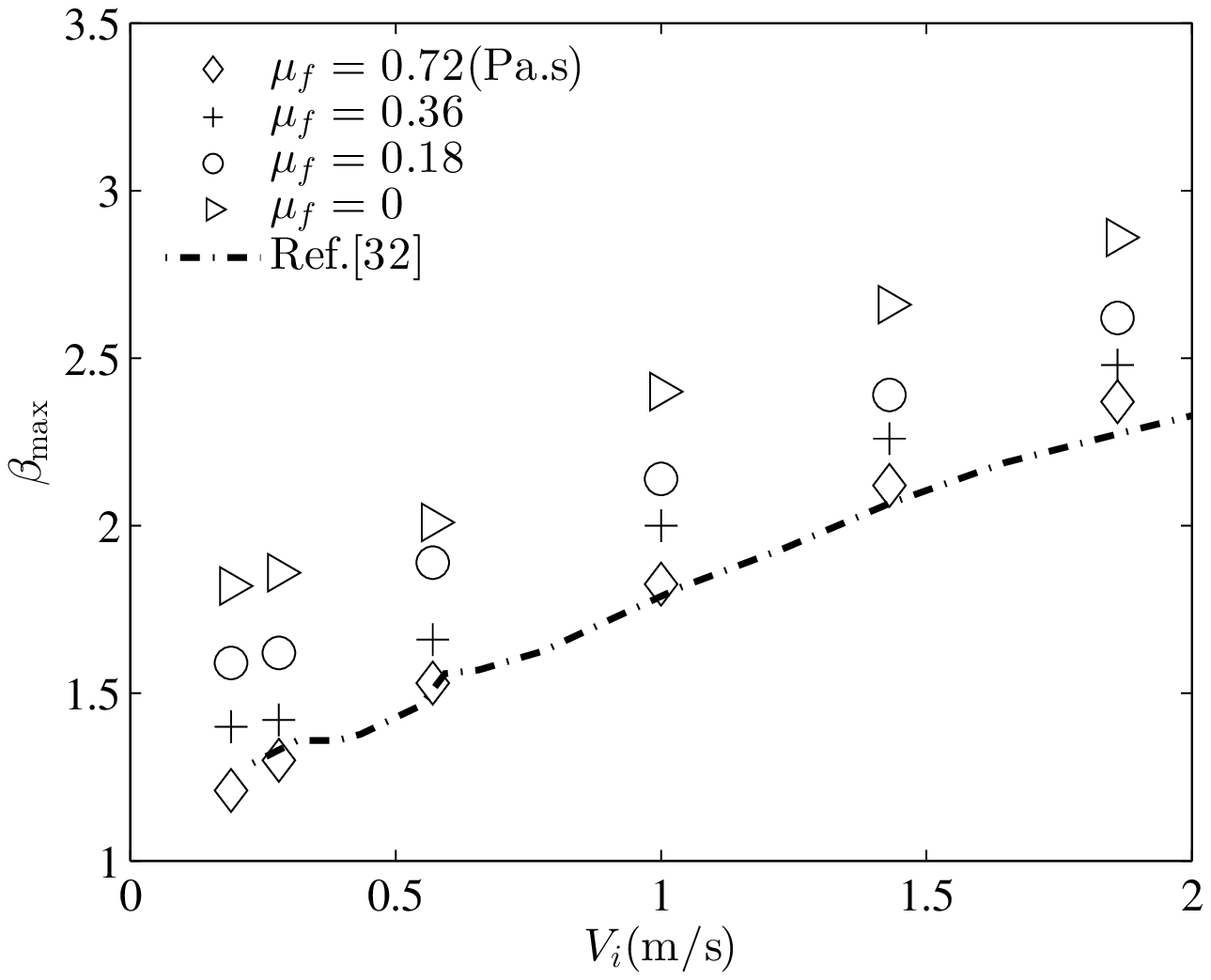}
     \put(-255,170){ $\left(a\right)$} \\
    \includegraphics[width = 0.55\textwidth,trim = 0cm 0cm 0cm 0cm, clip=true]{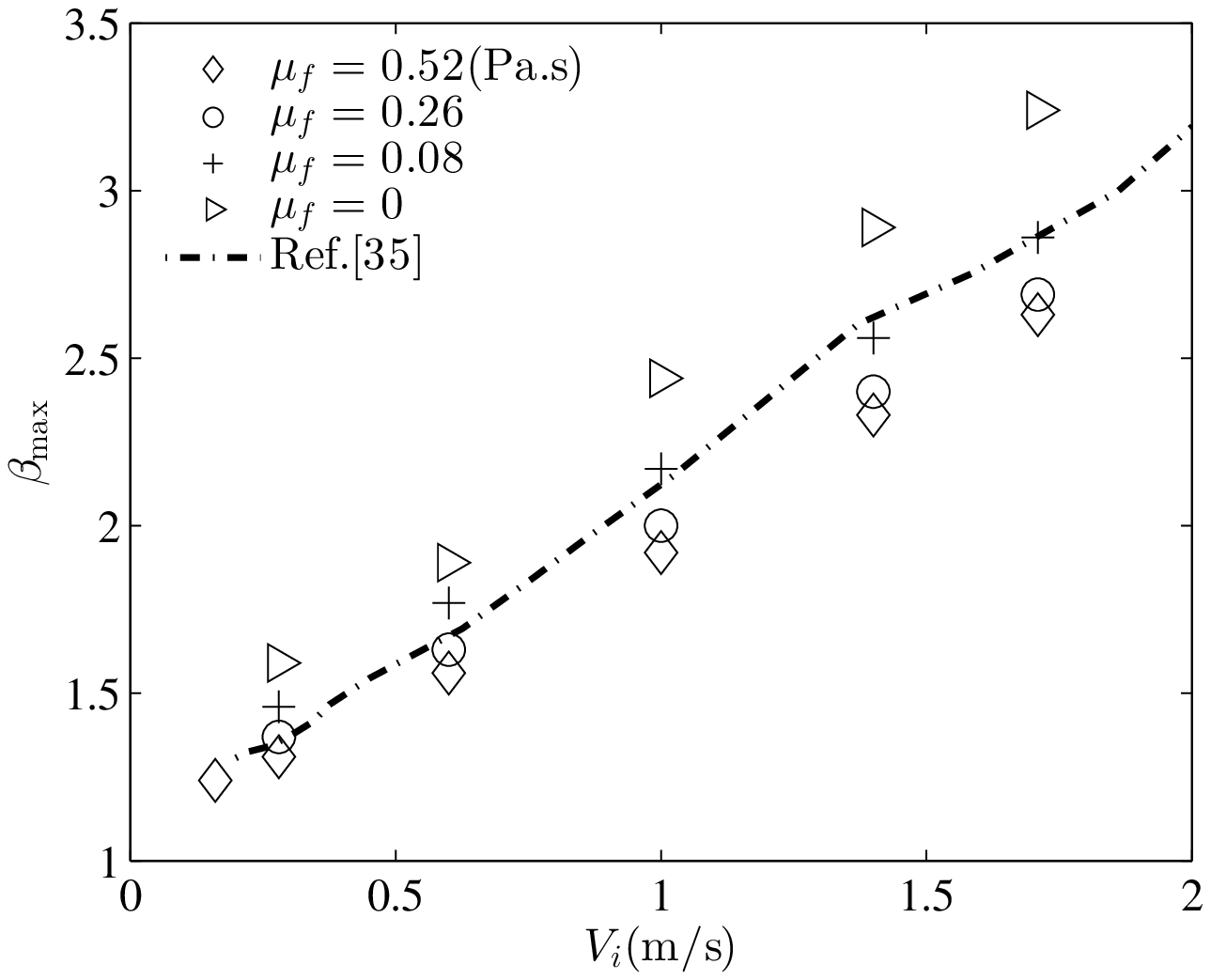}
     \put(-255,170){ $\left(b\right)$}
    \caption{\label{f4} The maximum spreading factor $\beta_\mathrm{max}=\max(\beta(t))$ as a function of impact speed $V_i$ and $\mu_f$. (a) Droplets with a glycerol-water mixture impacting onto a steel substrate, where the dashed line is interpolated experimental data \citep{lee2015universal} (their Fig.4). (b) Water droplets impacting onto an aluminum substrate, where the dashed line is interpolated  experimental data \citep{lee2016modeling} (their Fig.4(b)).}
    \end{figure}

It is clear that $\mu_f$ needs to be determined individually for each air-liquid-substrate combination. A too small value for $\mu_f$ causes an over-prediction of $\beta_\mathrm{max}$, while a too large value for $\mu_f$ causes an under-prediction of $\beta_\mathrm{max}$, see Fig.\ref{f4}. We want to highlight that the value for $\mu_f$ determined from a single experiment is also the best-fit for a range of impact speeds $V_i$ and shows that $\mu_f$ is not a function of $V_i$.
%
%makes the droplet spread too fast across the substrate and a too large value makes the spreading too slow an under prediction of $\beta_\mathrm{max}$, e.g., see curves of $\mu_f = 0.26\mathrm{Pa.s},0.52\mathrm{Pa.s}$ for water droplets on the aluminum substrate in Fig.\ref{f4}(b) and a too small value for $\mu_f$ makes the droplet spread too fast. In Fig.\ref{f4} we illustrate how $\mu_f$ influences $\beta_\mathrm{max}$ as a function of $V_\mathrm{i}$, where the best-fit $\mu_f$ is indeed the same as identified in Fig.\ref{f3}(a,b) for a single case, i.e., $\mu_f = 0.72 \mathrm{Pa.s}$ (Fig.\ref{f4}(a)) for glycerol-water and $\mu_f = 0.52\mathrm{Pa.s}$ for water on steel (Fig.\ref{f3}(c,d)).

\subsection{Droplet impact on textured substrates}
\begin{figure}[h]
     \centering
    %\begin{align}
     \includegraphics[width = 0.6\textwidth,trim = 0cm 0cm 0cm 0cm, clip=true]{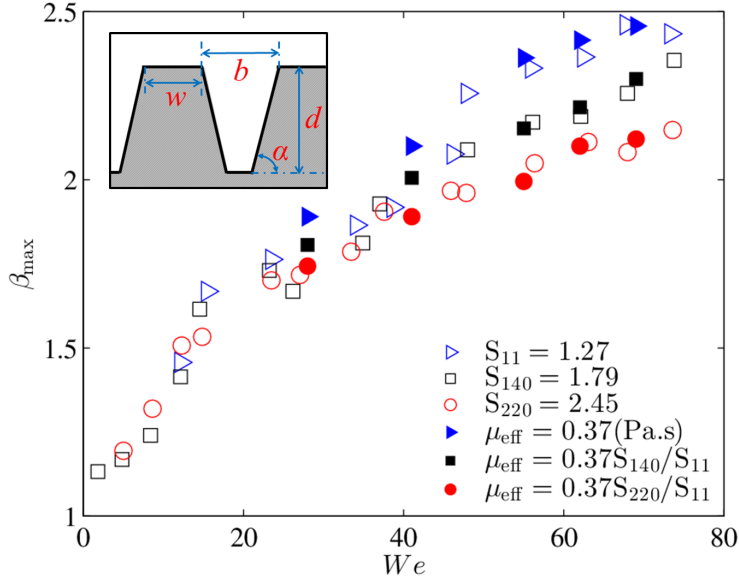}
      \caption{\label{f6} $\beta_\mathrm{max}$ for water droplets impacting onto micro-textured stainless steel substrates with grooves along one direction (see inset). The filled markers are numerical simulations with $\mu_\mathrm{eff} \sim S\mu_f$. The hollow markers are experimental results (Fig.3(b) in\cite{vaikuntanathan2016maximum}). The inset shows the geometric parameters ($w,b,d$ and $\alpha$) for the grooved substrate (Fig.1 in\cite{vaikuntanathan2016maximum}).}
         \end{figure}

Another parameter that can influence droplet spreading upon impact is the substrate topography\citep{vaikuntanathan2016maximum,robson2016asymmetries}. For spontaneous spreading of droplets ($V_i=0$) the friction factor $\mu_f$ has already been shown to rationalize spreading dynamics on rough substrates where the magnitude of $\mu_f$ depends on the substrate roughness factor ($S$). To test if our description of the contact line dynamics can provide a universal framework that can effectively bridge impact dynamics on smooth and rough substrates, we test the relation for the effective contact line friction parameter $\mu_{\mathrm{eff}} \sim S \mu_f$\citep{wang2015surface}, having already estimated the value for $\mu_f$ for the smooth substrate. The geometry of the textured substrate gives the following roughness factor,
\begin{equation}\label{eqs}
S = \frac{b+w-2d\mathrm{cos}(\alpha) + 2d/\mathrm{sin}(\alpha)}{b+w},
\end{equation}
which is the ratio of the real area and projected area of the substrate. $b,w,d,\alpha$ are geometric parameters describing the grooved substrate, see inset in Fig.\ref{f6}. If the contact line friction parameter $\mu_f$ is known for the corresponding flat substrate, the effective friction $\mu_{\mathrm{eff}}$ can easily be determined once the geometry of the micro-textured substrate is known.

We compare our simulations with experiments on substrates that have grooves along one direction with different aspect ratios. In Fig.\ref{f6} the effective contact line friction parameter $\mu_\mathrm{eff}$ is determined by matching the experimental data for the TS11 substrate\citep{vaikuntanathan2016maximum} with a roughness factor $S_{11}=1.27$, where we test the relation $\mu_{\mathrm{eff}} \sim S \mu_f$\citep{wang2015surface} for the substrates TS140 ($S_{140}=1.79$) and TS220 ($S_{220}=2.45$)\citep{vaikuntanathan2016maximum}. Although our assumption of axial-symmetry is slightly violated in the experiments, the linear relationship between $S$ and $\mu_\mathrm{eff}$ rationalizes $\beta_{max}$ for the spreading perpendicular to the grooves. This is also believed to be the primary cause for the difference in $\mu_f$ found for a water droplet impacting on the smooth ($\mu_f=0.08\mathrm{Pa.s}$) and textured ($\mu_f= 0.28\mathrm{Pa.s}$) stainless steel substrate.
%,, using the same approach domesticated by Fig.\ref{f4}. The values of $\mu_f$ of the other two surfaces made of the same material can be estimated as $\mu_{f\_\mathrm{T140}} =  S_\mathrm{140}/S_\mathrm{11} \mu_{f\_\mathrm{T11}}$, $\mu_{f\_\mathrm{T220}} =  S_\mathrm{220}/S_\mathrm{11} \mu_{f\_\mathrm{T11}}$, provided their geometric parameters\cite{vaikuntanathan2016maximum}.
%
%The groove depth $d$ tends to be larger from the TS11 surface to the TS220 surface, $S$ increases accordingly, i.e, the surface become rougher. An increase of effective contact line friction is expected and the droplet accordingly spreads less on the surface, shown in Fig.\ref{f6}. Using the predicted values of $\mu_f$ in our simulations, reasonable agreement on $\beta_\mathrm{max}$ with the experiment is reached. The experiment in the literature\citep{vaikuntanathan2016maximum} uses groove stripes while our layout is axisymmetric, thus Fig.\ref{f6} only presents their measurement conducted at the spreading end going cross the groove. The spreading along the stripes behaves quite similar to on a flat surface (their Fig.3) and results in a larger $\beta$, this further support that the grooves hinder the spreading.

\subsection{Energy budget}
\begin{figure}[h]
     \centering
    %\begin{align}
     \includegraphics[width = 0.65\textwidth,trim = 0cm 0cm 0cm 0cm, clip=true]{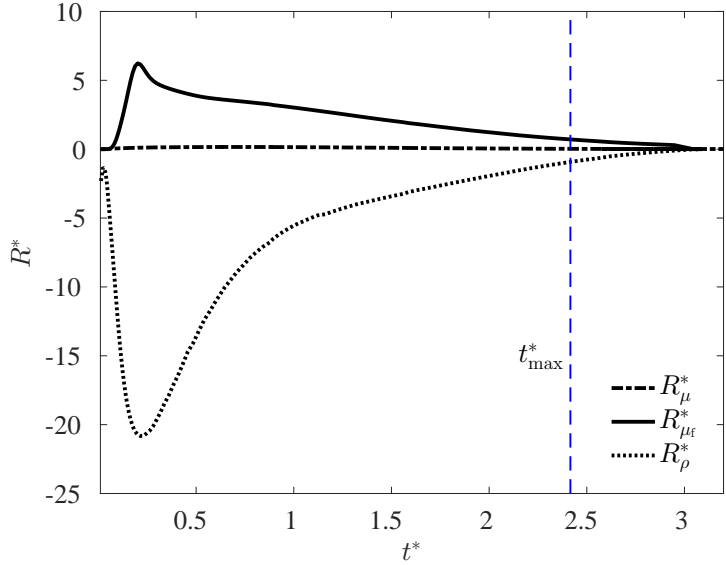}
      \caption{\label{f7} The non-dimensional rate of change of kinetic energy ($R_\mathrm{\rho}^* = \int_\mathrm{\Omega} \frac{1}{2} Ca \frac{1}{2}Re \frac{\partial ( \rho^*(C) \mathbf{u}^{*2} )} {\partial t^*} \mathrm{d}\Omega$), rate of viscous dissipation ($R_\mathrm{\mu}^* = \int_\mathrm{\Omega} Ca \mu^*(C) \left( \nabla \mathbf{u}^* + \nabla \mathbf{u}^{*\mathbf{T}} \right): \nabla \mathbf{u}^* \mathrm{d}\Omega$) and contact line dissipation ($R_\mathrm{\mu_f}^* = \int_\mathrm{\Gamma}  D_w  \left(\frac{\partial C}{\partial t^*} \right)^2 \mathrm{d}\Gamma$) for a glycerol-water droplet (10 mPa.s) impacting with; $V_i=1 \mathrm{m/s}, R = 0.92 \mathrm {mm}$ and $\mu_f = 0.72 \mathrm{Pa.s}$ i.e. $Re = 212, Ca = 0.15, We = 31.4$. $t_\mathrm{max}^* = 2.41$ is here the time in which the droplet is most deformed along r-direction $\beta_\mathrm{max} = 1.83$.}
     \end{figure}

Our results show that the local interface-wall contact line friction can affect the droplet impact dynamics, and we want next to determine its dissipative contribution and to compare it against the other primary contributions in the energy budget. To do this, we extract the different rates of energy and dissipations, where the principal contributions are; the rate of change of kinetic energy $R_\mathrm{\rho} = \frac{1}{2}\int_\mathrm{\Omega} \partial (\rho(C)\mathbf{u}^2)/ \partial t \mathrm{d}\Omega  $, the rate of viscous dissipation $R_\mu = \int_\mathrm{\Omega} \mu(C) \left( \nabla \mathbf{u} + \nabla \mathbf{u}^{\mathbf{T}} \right):\nabla \mathbf{u} \mathrm{d}\Omega$, and the rate of contact line dissipation $R_{\mu_f} = \int_\mathrm{\Gamma} \epsilon \mu_f (\partial C / \partial t)^2 \mathrm{d}\Gamma$. $\Omega$ is here the entire volume and $\Gamma$ is the substrate area. In the droplet impact dynamics we observe that at early times $t^*<0.25$ the magnitude of $R_\mathrm{\rho}^*$ decreases rapidly, while $R_\mathrm{\mu_f}^*$ on the other hand increases, see Fig.\ref{f7}. A minimum in $R_\mathrm{\rho}^*$ and a maximum in  $R_\mathrm{\mu_f}^*$ take place at $t^*\approx0.25$, whereas both slowly approach zero as the velocities decrease. Surprisingly, viscous dissipation appears to not play an important role in this regime as both $R_\mathrm{\mu_f}^*$ and $R_\mathrm{\rho}^*$ are much larger for $\beta(t) < \beta_\mathrm{max}$.
%$R_\mathrm{\rho},R_\mu$ and $R_{\mu_f} $ are the principal contributions to the rates of energy transfer in the dynamics, and the other contributions are not shown here as they are small in comparison.

\subsection{Scaling laws for $\beta_\mathrm{max}(\mu_f)$}
\begin{figure}[h]
    \centering
    \includegraphics[width = 0.65\textwidth,trim = 0cm 0cm 0cm 0cm, clip=true]{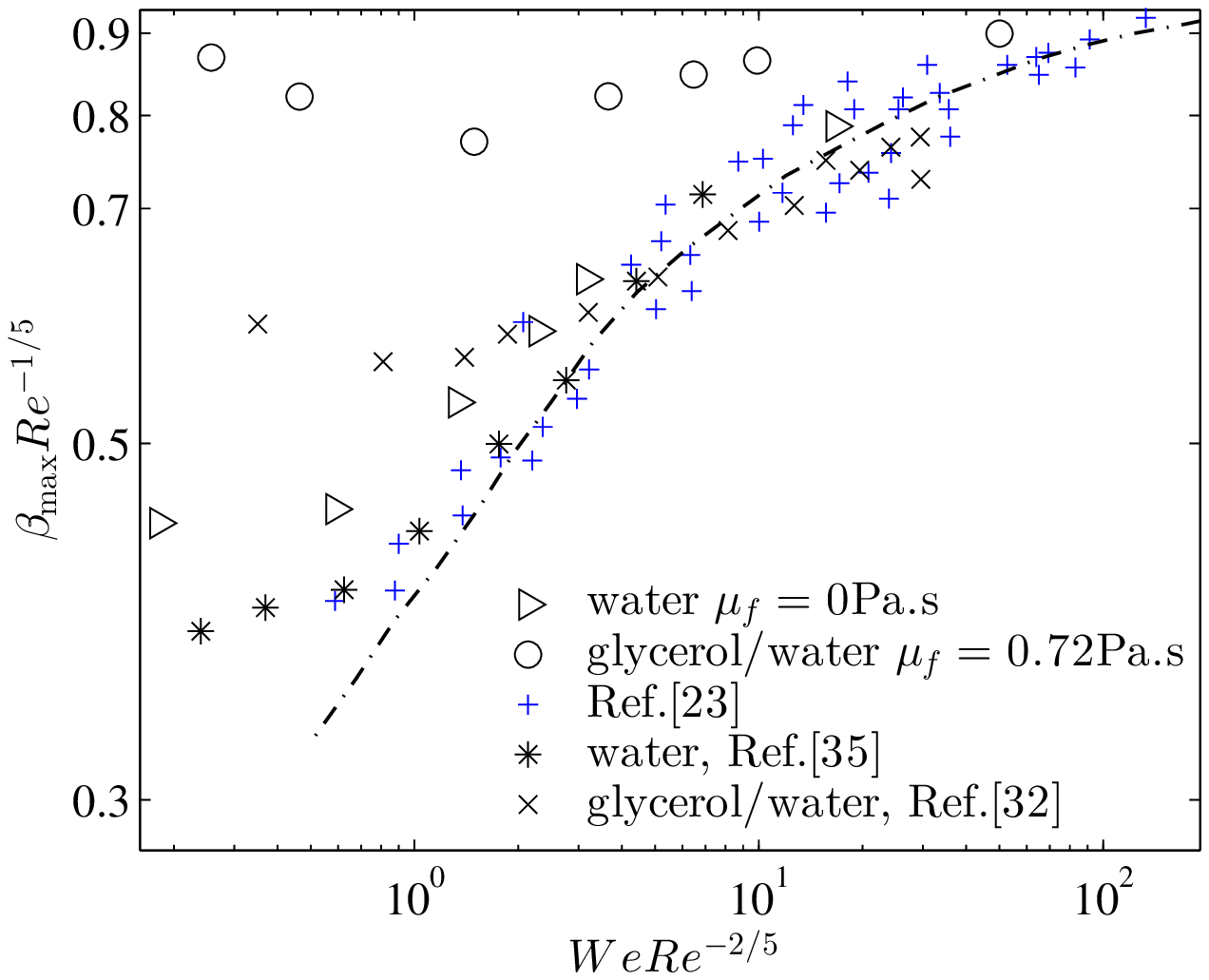}
    \put(-245,155){ $\left(a\right)$} \hspace{1.5mm}
    \includegraphics[width = 0.65\textwidth,trim = 0cm 0cm 0cm 0cm, clip=true]{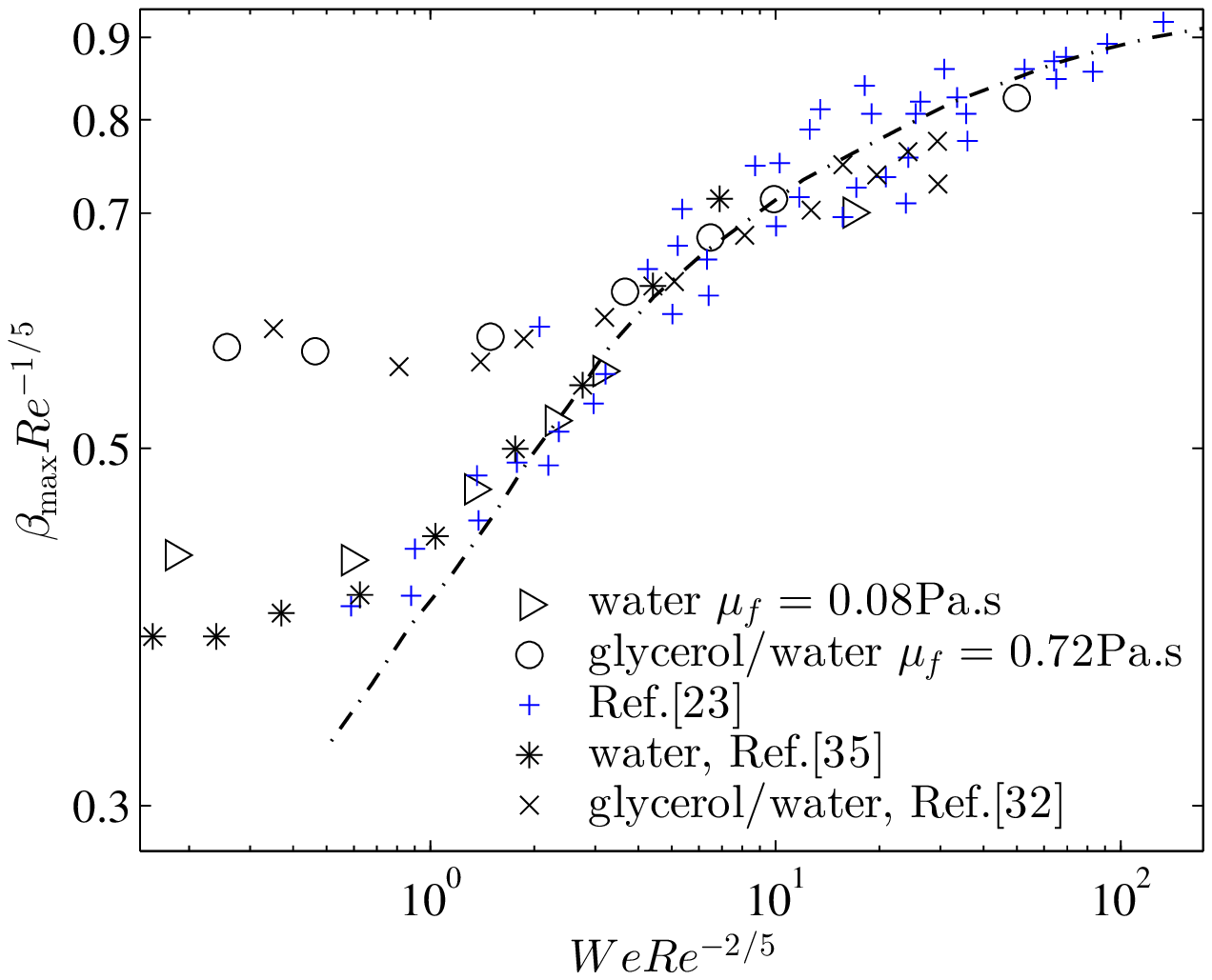}
    \put(-245,155){ $\left(b\right)$} \hspace{1.5mm}
    \caption{\label{f8}The maximum spreading factor $\beta_\mathrm{max}$ for different liquids, from experiments\cite{lee2015universal,laan2014maximum,lee2016modeling} (symbols; $+,\times,*$) and simulations (hollow markers). The dashed line is illustrating\cite{josserand2016drop} the scaling relation $\beta_\mathrm{max} = Re^{\frac{1}{5}}f(WeRe^{-\frac{2}{5}})$. (a) Comparing experiments\cite{lee2015universal,laan2014maximum,lee2016modeling} with numerical simulations for $\mu_f =0$. (b)Comparing experiments\cite{lee2015universal,laan2014maximum,lee2016modeling} with numerical simulations for $\mu_f$ as reported Table 2.}
    \end{figure}

Since our simulations fall into both the inertia-viscous and the inertia-capillary regime, we can further test if our numerical simulations are also consistent with existing scaling laws. We compare our numerical data for $\mu_f = 0$ (Fig.\ref{f8}a) and the measured $\mu_f $ (Fig.\ref{f8}b) with another set of independent experimental data\cite{laan2014maximum} for impacting droplets of different fluids. We use the scaling law for $\beta_\mathrm{max} = Re^{\frac{1}{5}}f(WeRe^{-\frac{2}{5}})$ that couples the inertia-viscous and inertia-capillary regime\citep{eggers2010drop,laan2014maximum} and illustrated by the line in Fig.\ref{f8}. We see in Fig.\ref{f8}(a) that $\mu_f=0$ creates results that deviate from the scaling law and the experiments. One exception is the simulations of water droplets, this is not surprising as the Reynolds number for water is ten times larger than for the glycerol-water mixture and inertial effects are therefore much more dominant. Including the effect of contact line dissipation by using the values for $\mu_f$ determined in Fig.2,4 makes the simulated data for $WeRe^{\frac{-2}{5}}>1$ fit well with the scaling law $\beta_\mathrm{max} = Re^{\frac{1}{5}}f(WeRe^{-\frac{2}{5}})$ and the experiments. However, it is clear that for the regime where the effect from $\mu_f$ is expected to be important i.e. $WeRe^{\frac{-2}{5}}<1$, both the experiments and the simulations deviate from the existing scaling law.

%We found that our simulation data of $\mu_f = 0.72\mathrm{Pa.s}$ for glycerol droplets on a steel substrate, and $\mu_f = 0.08\mathrm{Pa.s}$ for water droplets on a aluminum substrate collide with the scaling curve. However, both our simulation data and the experimental data are clearly off the scaling tendency near the lower limit, proving that this scaling still fails to predict the maximum spreading at very low impact speed, e.g., $V_\mathrm{i} < 0.25 \mathrm{m/s}$.  Meanwhile our results for water is less sensitive to the magnitude of $\mu_f$ because $Re$ for water is typically $10$ times larger than glycerol-water mixture(Table.\ref{tab1}) at the same $V_i$ thus the impact is dominant by inertia. The selectivity of glycerol-water mixture on $\mu_f$ indicates that the contact line dissipation is an important mechanism especially for a viscous impact, a new scaling can be derived considering $\mu_f$.

\begin{figure*}[h]
    \centering
    \includegraphics[width = 1\textwidth,trim = 0cm 0cm 0cm 0cm, clip=true]{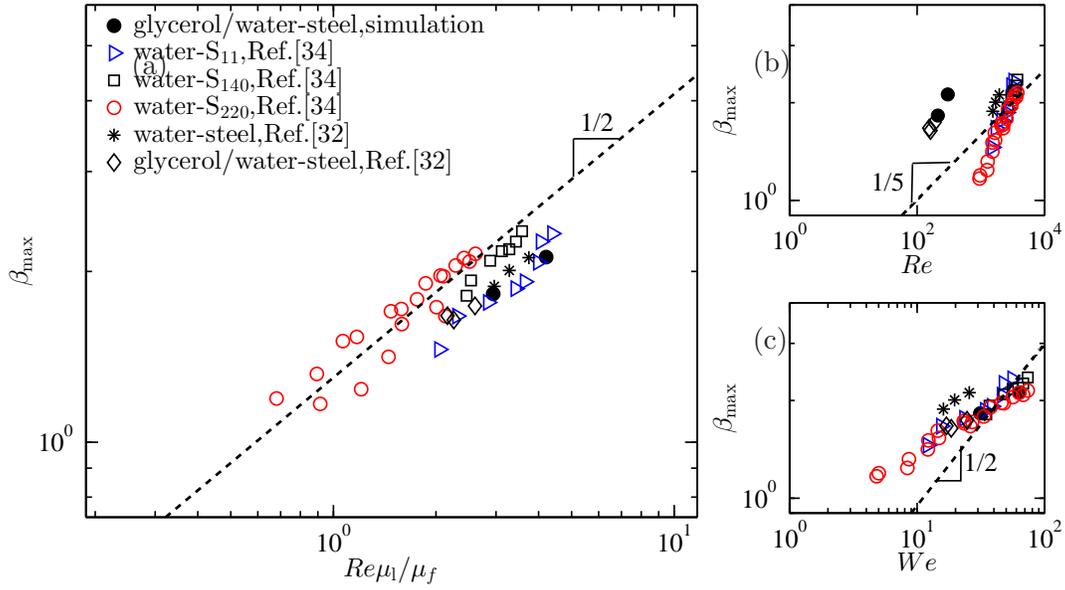}
    \put(-390,194){\small{(a)}}
    \put(-155,195){\small{(b)}}
    \put(-155,90){\small{(c)}}
    \caption{\label{f9} Scalings for $\beta_\mathrm{max}$ with data from droplet impact on different substrates where $\frac{\mu_f}{\mu_\mathrm{l}} > 5(\beta_\mathrm{max}^3 + \frac{1}{Ca})$. (a)$\beta_\mathrm{max}$ as a function of $\frac{\mu_\mathrm{l}}{\mu_f}Re$. (b)$\beta_\mathrm{max}$ as a function of $Re$. (c)  $\beta_\mathrm{max}$  as a function of $We$. Note that for the grooved substrates $\mathrm{S_{11},S_{140}}$ and $\mathrm{S_{220}}$, we use the measured valued for $\mu_\mathrm{eff}$.}
    \end{figure*}

To improve the scaling prediction for $\beta_\mathrm{max}$ we include the influence from the contact line dissipation into the approximation for the energy balance,
\begin{equation}\label{eq3.3}
\int_\Omega \frac{1}{2}\frac{\partial (\rho_\mathrm{l} \mathbf{u}^2)}{\partial t} \mathrm{d} \Omega \approx \int_\Omega \mu_\mathrm{l} \left( \nabla \mathbf{u} + \nabla \mathbf{u}^{\mathbf{T}} \right):\nabla \mathbf{u}  \mathrm{d}\Omega + \int_R \mu_fV_\mathrm{c}^2 \mathrm{d}R + \int_s \sigma \frac{\partial s}{\partial t} \mathrm{d}s
\end{equation}
with the contact line speed defined as $V_\mathrm{c} = \partial R(t) / \partial t$.
Note that the contact line dissipation is independent of the interface thickness and can be re-written\cite{carlson2011dissipation} as $\int_\mathrm{\Gamma} \epsilon \mu_f (\partial C / \partial t)^2 \mathrm{d}\Gamma=\int_{R_c} \mu_fV_\mathrm{c}^2 \mathrm{d}{R_c}$ with $R_c$ as the radial position of the contact line. $s$ is the droplet surface area and the last term on the right hand side of Eq.\ref{eq3.3} is the rate of change of droplet surface energy. Based on the approximated energy balance we assume the following scaling relations\cite{josserand2016drop} for a droplet that has spread to $R_\mathrm{max}$ and has a height $h$; $\mathbf{u}^2 \sim V_\mathrm{i}^2$, $\nabla \mathbf{u} \sim V_\mathrm{i}/h$, $t \sim R_\mathrm{max}/V_\mathrm{i}$, $V_\mathrm{c} \sim V_\mathrm{i}$, $\Omega \sim R^3$, $\Gamma \sim R_\mathrm{max}$, $s \sim R_\mathrm{max}^2$. In addition mass conservation of the incompressible droplet demands that its volume remains constant and that $R^3\sim hR_\mathrm{max}^2$. By introducing these scaling relations into Eq.\ref{eq3.3} we get the following expressions:

 \begin{equation}\label{eq3.4}
 \begin{split}
  & \int_\Omega \frac{\partial (\rho_\mathrm{l} \mathbf{u}^2)}{\partial t} \mathrm{d} \Omega \sim \frac{\rho_\mathrm{l}V_\mathrm{i}^3R^3}{R_\mathrm{max}}\\
  & \int_\Omega \mu_\mathrm{l} \left( \nabla \mathbf{u} + \nabla \mathbf{u}^{\mathbf{T}} \right):\nabla \mathbf{u}  \mathrm{d}\Omega \sim \frac{\mu_\mathrm{l} V_\mathrm{i}^2R_\mathrm{max}^4}{R^3}\\
  & \int_\Gamma \mu_fV_\mathrm{c}^2 \mathrm{d}\Gamma \sim \mu_f V_\mathrm{i}^2 R_\mathrm{max} \\
  & \int_s \sigma \frac{\partial s}{\partial t} \mathrm{d}s \sim \sigma V_\mathrm{i} R_\mathrm{max}.
 \end{split}
 \end{equation}
Now substituting these scaling relations into Eq.\eqref{eq3.3} and rearranging the terms give $\beta_\mathrm{max}$ as a function of the contact line friction parameter $\mu_f$,
\begin{equation}\label{eq3.6}
Re= \beta_\mathrm{max}^2\left( \frac{\mu_f}{\mu_\mathrm{l}} + \frac{1}{Ca} + \beta_\mathrm{max}^3  \right).
\end{equation}
Three separate regimes appear, where we immediately see that the two limits $\frac{\mu_f}{\mu_\mathrm{l}} + \frac{1}{Ca} \ll \beta_\mathrm{max}^3$ ($\beta_\mathrm{max}\sim Re^{\frac{1}{5}}$) and $\frac{\mu_f}{\mu_\mathrm{l}}  + \beta_\mathrm{max}^3\ll\frac{1}{Ca}$ ($\beta_\mathrm{max}\sim We^{\frac{1}{2}}$) recovers the classical scaling laws. However, we identify a new regime for small and intermediate impact speeds i.e. $\frac{1}{Ca} + \beta_\mathrm{max}^3\ll \frac{\mu_f}{\mu_\mathrm{l}}$ with $\beta_\mathrm{max}\sim (\mu_\mathrm{l}Re/\mu_f)^{\frac{1}{2}}$.To test this new scaling law we plot the data for $\mu_f /\mu_\mathrm{l} > 5(\beta_\mathrm{max}^3 + 1/Ca)$ from the simulations and experiments, which follows $\beta_\mathrm{max}\sim (Re \mu_\mathrm{l}/ \mu_f)^{\frac{1}{2}}$, see Fig.\ref{f9}(a), instead of $\beta_\mathrm{max}\sim We^{\frac{1}{2}}$ or $\beta_\mathrm{max}\sim Re^{\frac{1}{5}}$, see Fig.\ref{f9}(b,c).
%Indeed our numerical data follows this scaling as $\beta_\mathrm{max} = 1.38 \sqrt{Re\mu_\mathrm{l}/\mu_f}$, but scatters in the scaling $Re^{1/5}$, see Fig.\ref{f9}.

%It is clear that if $\beta_\mathrm{max}^3\mu_\mathrm{l}/\mu_f \gg 1$ and $\mu_\mathrm{l}/(Ca\mu_f) \sim O(1)$, Eq.\eqref{eq3.6} gives that $\beta_\mathrm{max}$ should scale with $Re^{1/5}$. In the limit of $\mu_\mathrm{l}/(Ca\mu_f) \gg 1$, while $\beta_\mathrm{max}^3\mu_\mathrm{l}/\mu_f \sim 1$, $\beta_\mathrm{max}$ is predicted to scale with $We^{1/2}$. For $\beta_\mathrm{max}^3\mu_\mathrm{l}/\mu_f \ll 1$ and $\mu_\mathrm{l}/(Ca\mu_f) \ll 1$, $\beta_\mathrm{max}$ is predicted to scale with $(Re\mu_\mathrm{l}/\mu_f)^{1/2}$. Both the scaling laws $Re^{1/5}$ and $We^{1/2}$ are well known, but the regime for low impact speeds, i.e., $(Re\mu_\mathrm{l}/\mu_f)^{1/2}$ has not been demonstrated previously.

\section{Conclusions}
We have investigated the dynamics of droplets impacting onto solid substrates as a function of their viscosity, substrate wettability and substrate topography by deploying numerical simulations. By assuming linear response through a Stokes-like drag at the contact line, our simulations rationalize experimental observations for droplet impact on both smooth and rough substrates. Our results highlight that at low impact speeds the dissipation at the contact line needs to be included to predict the droplet spreading dynamics. We propose a scaling relation for this regime that is dominated by contact line dissipation $\beta_\mathrm{max}\sim(\frac{\mu_\mathrm{l}}{\mu_f}Re)^{\frac{1}{2}}$, complementing the classical scaling laws for $\beta_\mathrm{max}$ i.e. $\beta_\mathrm{max}\sim We^{\frac{1}{2}}$ and $\beta_\mathrm{max}\sim Re^{\frac{1}{5}}$ that are also identified in the numerical simulations. Moreover, our simulations highlight the link between substrate roughness and the effective contact line friction factor $\mu_\mathrm{eff}$ that can provide a unifying framework to describe droplet impact dynamics.


\end{document}